\documentclass[11pt,letterpaper]{article}
\usepackage[hyperref]{tto2022}
\usepackage{times}
\usepackage{latexsym}
\usepackage{graphicx}
\usepackage{url}
\usepackage{subcaption}
\usepackage{float}
\usepackage{amsmath}
\usepackage{cleveref}
\usepackage{censor}

\aclfinalcopy 

\title{The role of online attention in the supply of disinformation in Wikipedia}

\author{Anis Elebiary$^\dag{}$ \and Giovanni Luca Ciampaglia$^{\dag{}\,\ddag{}}$ \\
  ${}^\dag{}$ University of South Florida, Tampa, FL \\
  ${}^\ddag{}$ University of Maryland, College Park, MD \\
  \texttt{anis@usf.edu},\quad{} \texttt{gciampag@umd.edu}}

\date{}

\begin{document}

\maketitle

\begin{abstract}
  Wikipedia and many User-Generated Content (UGC) communities are known for
  producing reliable, quality content, but also for being vulnerable to false
  or misleading information. Previous work has shown that many hoaxes on
  Wikipedia go undetected for extended periods of time. But little is known
  about the creation of intentionally false or misleading information online.
  Does collective attention toward a topic increase the likelihood it will
  spawn disinformation? Here, we measure the relationship between allocation of
  attention and the production of hoax articles on the English Wikipedia.
  Analysis of traffic logs reveals that, compared to legitimate articles
  created on the same day, hoaxes tend to be more associated with traffic
  spikes preceding their creation. This is consistent with the idea that the
  supply of false or misleading information on a topic is driven by the
  attention it receives. These findings improve our comprehension of the
  determinants of disinformation in UGC communities and could help promote the
  integrity of knowledge on Wikipedia. \end{abstract}

\section{Introduction}

In recent years several Internet websites have become the hubs for communities
where users can produce, consume, and disseminate content without central
oversight. Examples of these \emph{user-generated content} (UGC) websites
include major social media platforms, like Facebook or Twitter, or global
online knowledge production communities like Wikipedia, which is known as a
model for the production of vast reliable, high-quality
knowledge~\cite{yasseri2021crowdsourcing}.

However, a negative consequence of the popularity of UGC websites is that their
low barriers to access, combined with the lack of supervision from experts or
other gatekeepers, results in the proliferation of false or misleading
information on the Web as a whole~\cite{wardle2017information,
lazer2018science}.  

False or misleading content often spreads on social networking
platforms~\cite{amoruso2020contrasting,castillo2011information,zareie2021minimizing,grinberg2019fake,guess2019less,guess2020exposure,allcott2017social},
but there are growing concerns that other UGC communities like Wikipedia may be
vulnerable to these threats too~\cite{diego2019online}. This is especially
worrisome since Wikipedia is one of top most visited internet
websites~\cite{similarweb2022websites} and a popular source of
knowledge~\cite{okoli2014wikipedia}. 
Wikipedia contains over 50 million articles in more than 300 languages; in
February 2022, the English language edition of Wikipedia alone received 781M
visits (from unique devices) and was edited over 5M
times~\cite{wikimedia2022size,wikimedia2022wikimedia}. Hence, preserving the
integrity of Wikipedia is of paramount importance for the Web as a
whole~\cite{diego2019online}. 

There are many potential threats to the integrity of knowledge in
Wikipedia~\cite{diego2019online}. One common threat comes from vandalism, which
is ``a deliberate attempt to compromise the integrity of the encyclopedia,
often through the insertion of obscenities, insults, nonsense or crude humour,
or by page blanking''~\cite{wikimedia2022vandalism}. 

Vandalism, however, is not the only threat to the integrity of Wikipedia's
content. Whereas vandalism focuses on defacing existing entries, there exists
evidence showing that Wikipedia is also targeted by \emph{hoaxes}, whose aim is
to create whole new entries about fake, fictitious topics. An example of a
famous Wikipedia hoax is the entry \emph{Jar'Edo Wens}, a fake Australian
Aboriginal deity, which went undetected for almost 10 years before being
debunked and deleted~\cite{dewey2015story}. But hoaxes remain a threat to
Wikipedia's content integrity to this day. Recently, one of the largest such
incidents the platform has ever seen has been discovered on the Chinese
Wikipedia: a user named \emph{Zhemao} wrote 206 fake entries, starting from
2019 until 2022, about Russia's history in the Middle
Ages~\cite{moon2022chinese}.

Hoaxes are thus not to be confused with vandalism; although vandalism is a much
bigger threat in scope and size compared to hoax articles, hoaxes constitute a
more subtle threat, which has received less attention compared to vandalism.  

A crucial question that remains unresolved is what drives the creation of
hoaxes on Wikipedia. Because their original authors are aware that these
articles are false, hoax articles are different from mere
\emph{misinformation}, but should rather be considered instances of
\emph{disinformation}~\cite{wardle2017information,lazer2018science}. As such,
understanding the factors that determine the supply of hoaxes on Wikipedia
could shed light on disinformation in general, including broader threats to the
integrity of the Web, like state-sponsored propaganda~\cite{king2017chinese,
zannettou2019disinformation,golovchenko2020cross} and conspiracy
theories~\cite{starbird2017examining}.

To bridge this gap, in this paper, we study the role of \emph{online
attention}, in the form of individual page views, in the supply of
disinformation in Wikipedia. The idea of an economy of attention was first
introduced by \citet{simon1971designing}, who observed that human attention is
a limited resource that needs to be allocated~\cite{goldhaber1997attention}.
Here, to quantify the flow of collective attention to individual topics of
knowledge, we take advantage of the unique Wikipedia traffic dataset and API.
Specifically, in this work we seek to answer the following questions:
\begin{enumerate}
\item[Q1.] Does online attention toward a topic increase the likelihood of
disinformation being created about it?
\item[Q2.] Operationally, is there a relationship between traffic to Wikipedia
and the production of hoax articles?
\end{enumerate}

To answer these questions, we collected a list of known hoax
articles~\cite{wikimedia2022list} along with their creation timestamps and
content. To control for potential confounding factors in the distribution of
traffic to Wikipedia over time, for each hoax, we considered a cohort
consisting of all the legitimate (i.e.~non-hoax) Wikipedia articles that were
created on the same day as the hoax. Similar to
\citet{kumar2016disinformation}, we find that hoaxes differ from legitimate
articles in key appearance features, but do not strongly differ in the number
of hyperlinks they contain. Next, for each article (either hoax or non-hoax),
we parsed its content and extracted all the out-links, i.e. its neighbors in
the Wikipedia hyperlink network. The presence of a link between two Wikipedia
entries is an indication that they are semantically related. Therefore, traffic
to these neighbors gives us a rough measure of the level of online attention to
a topic \emph{before} a new piece of information (in this case an entry in the
encyclopedia) is created.

Finally, we measure the relative change of traffic in the 7-day period before
and after the creation of a hoax and compare this change to that of the
articles in its cohort. To preview our results, we find that, on average,
online attention tends to precede the creation of hoaxes more than it does for
legitimate articles. This observation is consistent with the idea that the
supply of false and misleading information on a topic is driven by the
attention it receives.

In the rest of the paper we discuss related work (\Cref{sec:related_work}), and
then describe our methodology (\Cref{sec:data_and_methods}): the details of the
data collection process, the comparison between features of hoaxes and
legitimate articles, and the pre-processing of the Wikipedia traffic data.
\Cref{sec:results} discusses the techniques used to quantify online attention
and its relationship to the hoax creation, and the statistical procedures
performed to asses the results. Finally, \cref{sec:discussion_and_future_work}
summarizes our findings and future directions.

All code and data needed to replicate the findings of this study are available
on Github at
\href{https://github.com/CSDL-USF/wikihoaxes}{github.com/CSDL-USF/wikihoaxes}.

\section{Related Work}
\label{sec:related_work}

Over the years Wikipedia has developed an impressive array of socio-technical
solutions to ensure the quality of its content. Early work on Wikipedia has
shown that most acts of vandalism are repaired manually by the crowd of
contributors within a matter of minutes~\cite{viegas2004studying}. In addition
to human interventions, automated tools like ClueBot~NG play a crucial role in
keeping the encyclopedic entries clear from
damage~\cite{geiger2013levee,halfaker2012bots}. On top of these methods, there
exist other preventive measures such as patrolling recent changes, creating
watchlists, blocking frequent vandalism creators, and using editorial filters.
Finally, multiple research attempts have been conducted to aid in both the
manual and the automatic detection of
vandalism~\cite{potthast2008automatic,adler2010detecting,smets2008automatic,harpalani2011language}. 

Despite this wealth of work, little is known about Wikipedia hoaxes.
\citet{kumar2016disinformation} collected a sample of known hoaxes from
resources compiled by the Wikipedia community, and studied their longevity,
along with other characteristics. They found that one in a hundred hoaxes
remain undetected for more than a year, with 92\% of the cases detected within
the first day. They also observed that, although only 1\% of all hoaxes remain
undetected for more than a year, those that stay undetected have a higher
chance over time of remaining so. Finally, they showed that, on average, hoaxes
have a lower density of internal links and receive less traffic than legitimate
(i.e.,~non-hoax) articles~\cite{kumar2016disinformation}.

Traffic to Wikipedia has been used before to study collective attention.
\citet{garcia2017memory} studied the patterns of attention to Wikipedia in the
wake of airplane crashes. They found that the traffic to entries about
\emph{previous} airplane crashes was greater than that of the current crash,
i.e. the one that triggered the attention surge~\cite{garcia2017memory}. 
\citet{ciampaglia2015production} studied traffic patterns during the creation
of new Wikipedia entries (i.e.,~not just hoaxes) and observed that the creation
of new information about a topic is preceded by spikes of attention toward it,
as measured by traffic to neighboring entries~\cite{ciampaglia2015production}.
This is consistent with a model in which the demand for information on a topic
drives the supply of novel information about it. Consequently, measuring
traffic to Wikipedia entries can help us get a step closer to understanding why
and when hoaxes are more likely to be produced. 

\section{Data and Methods}
\label{sec:data_and_methods}

We first describe how the dataset of hoaxes was collected and the process of
building the cohort of each hoax. 

\subsection{Data Collection}
\label{sec:data_collection}

Prior work has relied on a broad definition of `hoaxes' that leverages the `New
Page Patrol' (or NPP) process~\cite{kumar2016disinformation}. Unfortunately,
access to these data was not public due to the nature of the NPP process.
Therefore, in the present work we relied on a smaller, public list documenting
known hoaxes discovered by Wikipedia editors outside of the NPP
process~\cite{wikimedia2022list}. To be included in this list, a discovered
hoax must meet either of the following two characteristics: (\emph{i}\,) they
have gone undetected for more than a month after
patrolling~\cite{kumar2016disinformation}, or (\emph{ii}\,) they were discussed
by reliable media sources. 

\begin{figure*}[tbp]
  \centering
  \includegraphics[width=\columnwidth]{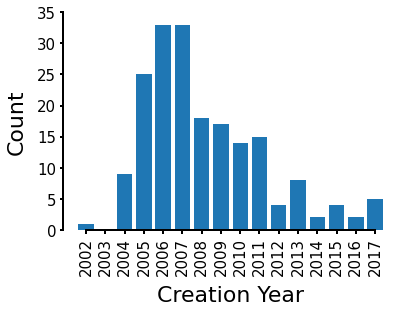}
  \includegraphics[width=\columnwidth]{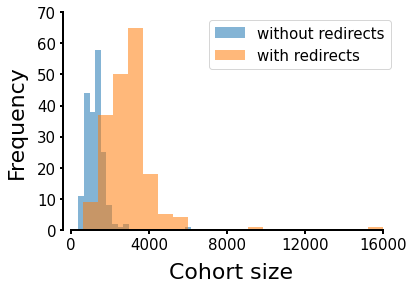}
  \caption{Left: Hoaxes detected in the English Wikipedia. Right: Cohort size
  distribution for hoaxes in our dataset before (solid blue) and after (solid
  orange) resolving redirects.} 
  \label{fig:creation_years_cohort_size}
\end{figure*}

To collect this list, we queried the Wikipedia API using the `prefix search'
endpoint~\cite{mediawiki2022main} to collect the titles of the hoaxes residing
in the administrative list maintained by Wikimedia under the prefix 'List of
Hoaxes on Wikipedia'. The total number of titles retrieved was $N_h=190$. We
then used the Toolforge~\cite{wikitech2021toolforge} to query the database
replica of the English Wikipedia for the creation date of each hoax article,
defined as the timestamp of the first revision recorded in the database.
\Cref{fig:creation_years_cohort_size}~(left) shows a summary of the number of
hoaxes created over time, with the majority of hoaxes appearing in the period
2005--2007, and a decline starting in 2008. This observed behavior can be in
part explained by the fact that the Wikipedia community started patrolling new
pages in November of 2007~\cite{kumar2016disinformation,wikimedia2022new} and
is also consistent with the well-known peak of activity of the English
Wikipedia community~\cite{halfaker2013rise}. 

Finally, to build the cohort of each hoax, we queried the Wikipedia database
replica for all legitimate articles created on the same day. Since Wikipedia
entries are often accessible through different titles, in collecting the
cohort, we resolved all redirects created the same day as the hoax. Treating
these redirects as separate entries would inflate the cohort size and could
skew traffic statistics used later for estimating the level of online
attention. \Cref{fig:creation_years_cohort_size}~(right) shows the effect that
redirects have on the size of each cohort. In some cases, failing to account
for redirects can increase the size of cohorts to up to $16,000$ articles. 

\subsection{Appearance Characteristics Analysis}
\label{sec:appearance_characteristics_analysis}

To understand the differences between each hoax and its cohort members, we
analyzed their appearance features, inspired by the work of
\citet{kumar2016disinformation} who, in addition to appearance features,
studied network, support, and editor features for both hoax and legitimate
articles~\cite{kumar2016disinformation}. We considered the following four
features: (\emph{i})~the \emph{plain text length} is the number of words in an
article after removing all the wiki markup; (\emph{ii})~the \emph{ratio of
plain to markup text} is the number of words obtained after removing all markup
in the body of the article, divided by the number of words before removal;
(\emph{iii})~the \emph{density of wiki-links} is the number of wiki-links per
100 words, counted before markup removal; and, finally, (\emph{iv})~the
\emph{density of external links} is defined similarly as the density of
wiki-links, but for links to external Web pages. 

To be able to calculate these features for each hoax and its cohort, we
consulted the API to extract their plain text using the \emph{TextExtracts}
extension instead~\cite{mediawiki2022text}. For the wiki markup we used the
revisions API~\cite{mediawiki2022revisions}. A regular expression was used to
count the number of words in plain and markup text. Finally, to find the wiki
and external links within each article we used
\emph{wikitextparser}~\cite{5j92022github}. 

\begin{figure}[tbp]
    \begin{subfigure}{0.23\textwidth}
        \includegraphics[width=\linewidth]{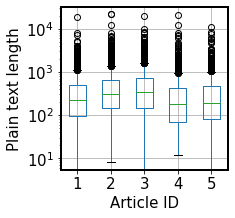}
    \end{subfigure}
    \begin{subfigure}{0.23\textwidth}
        \includegraphics[width=\linewidth]{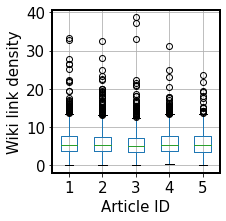}
    \end{subfigure}
    \begin{subfigure}{0.23\textwidth}
        \includegraphics[width=\linewidth]{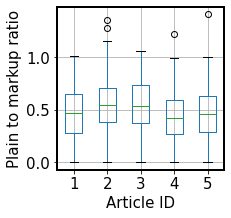}
    \end{subfigure}
    \begin{subfigure}{0.23\textwidth}
        \includegraphics[width=\linewidth]{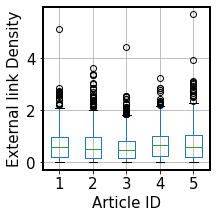}
    \end{subfigure}
    \caption{Distribution of appearance features for five manually sampled
    cohorts in our data.}
    \label{fig:outliers}
\end{figure}

Aside from the plain text to markup ratio, the chosen appearance features have
very skewed distributions. To illustrate this point, \cref{fig:outliers} shows
the distribution of each score for five manually sampled cohorts in our data.
For the plain text length, \cref{fig:outliers} shows that the median is between
$100$ and $1,000$ words, yet there exist articles that reach and even exceed
$10,000$ words. The same case persists in the wiki-link density --- the median
is under $10$ links per $100$ words, however some articles have up to $40$
links, and similar for the other two features.

Thus, after collecting all the four features, we computed the modified
$z$-score $z'$ to compare different hoaxes together:
\begin{equation}
z' = \frac{x - \tilde{x}}{\mathrm{MAD}} \\
\label{eq:z-score}
\end{equation}

Where $x$ is a feature measured on a hoax, $\tilde{x}$ the median value of the
feature on the non-hoaxes, and $\mathrm{MAD}$ the median absolute deviation of
$x$ with respect to $\tilde{x}$. We chose to use $z'$ instead of the regular
$z$-score since it is more resilient to outliers~\cite{iglewicz1993how}. 

\subsection{Analyzing Wikipedia Traffic Data}
\label{sec:analyzing_wikipedia_traffic_data}

To analyze the traffic that the articles in our dataset receive, we used a
dataset on traffic compiled by the Wikimedia
foundation~\cite{wikimedia2022page}. The Wikimedia Foundation has published two
main traffic datasets: the earlier \emph{pagecounts-raw} (Dec. 2007--Aug.
2015), and the more recent \emph{pageviews} (started Jul. 2015). Since most of
the hoaxes in our dataset were created in the period between 2005 and 2011, we
have decided to use the older pagecounts-raw data. This dataset contains the
count of non-unique HTTP requests made for each article in an hourly time
frame, collected by the proxy server of
Wikipedia~\cite{ciampaglia2015production}, along with request title and
additional metadata. We pre-processed pagecounts-raw to resolve redirects,
filter unwanted entries, and clean illegal titles. 

Pre-processing the data was performed over the following three steps. First,
the raw data was filtered. The filtration process selected only entries related
to the English Wikipedia project while removing all pages from namespaces other
than the `main' MediaWiki namespace. Second, the filtered data was cleaned from
illegal titles. Illegal titles were discarded by removing characters which are
not allowed in Wikipedia page titles~\cite[cf.~`Page
Restrictions']{wikimedia2022help}. The hashtag sign `\#' is considered illegal
only if it is the first character in a title; otherwise it indicates a
subsection within a page. Hence, a title including `\#' is discarded only in
the former case. In addition to removing illegal characters, we decoded common
URL-encoded characters (e.g. `\%20') and replaced any space with an underscore
character. Third, to resolve redirects, the Toolforge was consulted to extract
all the redirects within the main namespace of the English Wikipedia. The
result was a cleaned and filtered hourly dataset of the view count for pages
within the main namespace of the English Wikipedia.

\begin{figure*}[tbp]
    \begin{subfigure}{0.245\textwidth}
        \includegraphics[width=\linewidth]{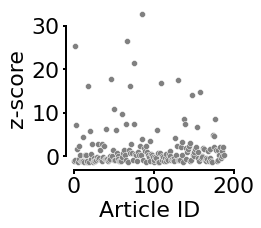}
        \caption{Plain text length}
        \label{fig:z-scores_plain}
    \end{subfigure}
    \begin{subfigure}{0.245\textwidth}
        \includegraphics[width=\linewidth]{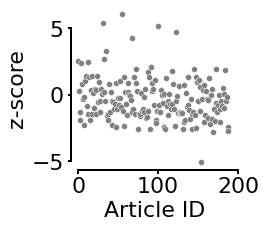}
        \caption{Wiki-link density}
        \label{fig:z-scores_wiki-link}
    \end{subfigure}
    \begin{subfigure}{0.245\textwidth}
        \includegraphics[width=\linewidth]{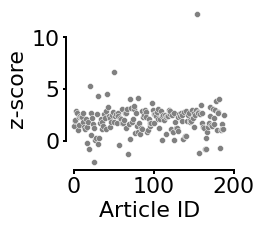}
        \caption{Plain to markup text ratio}
        \label{fig:z-scores_ratio}
    \end{subfigure}
    \begin{subfigure}{0.245\textwidth}
        \includegraphics[width=\linewidth]{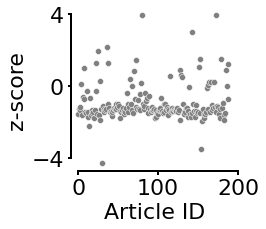}
        \caption{External link density}
        \label{fig:z-scores_ext-link}
    \end{subfigure}
    \caption{Modified $z$-scores for all hoaxes in our sample relative to non-hoax articles in their cohorts for the four appearance features we considered. Hoaxes tend to have similar or slightly smaller count of plain text words (however with several higher-count outliers), lower external link density, higher plain to markup text ratio, and similar wiki-link density.}
\end{figure*}

\section{Results}
\label{sec:results}

\subsection{Appearance Features}

We start by analyzing the appearance features of hoaxes relative to the
non-hoaxes in their cohort. \Cref{fig:z-scores_plain} shows that most hoaxes
have either similar or slightly smaller plain text length compared to that of
their cohorts. We also observe the presence of several outliers, indicating
that a subset of hoaxes in our sample tends to have unusually higher word
counts. This is consistent with the results of \citet{kumar2016disinformation},
who observed that `successful' hoaxes (i.e., that have gone undetected for at
least 1 month) have a median plain text length of $134$ words --- almost twice
as large as that of legitimate articles. However, the analysis of
\citet{kumar2016disinformation} differs from ours in multiple ways. First, as
already mentioned, they used a different, larger set of hoaxes collected as
part of Wikipedia's regular NPP process. Second, they used a matching procedure
to compare each hoax to only one legitimate article created on the same day.
They also considered other types of articles, such as wrongly flagged articles
and failed hoaxes. Another potential differentiating factor is the method of
extraction for the plain text, markup content, and links for each page, which
might contribute to not obtaining exactly the same results.

\Cref{fig:z-scores_wiki-link} shows that hoaxes tend to have a similar density
of wiki-links when compared to non-hoaxes. This is important, since to quantify
online attention toward a topic we compute the volume of traffic to the
wiki-link neighbors of an article. Thus, in the following analysis on traffic,
we can safely exclude potential confounding factors due to different linking
patterns between hoaxes and non-hoaxes. 

\Cref{fig:z-scores_ratio,fig:z-scores_ext-link} show the distributions of the
ratio of plain to markup text and of external link density, respectively. Aside
from a few outliers, hoaxes almost always contain more plain text than markup
text, compared with non-hoaxes. This is also consistent with the findings of
\cite{kumar2016disinformation}, who observed that, on average, 58\% of a
legitimate article, 71\% of a successful hoax, and 92\% of a failed hoax is
just plain text. 

In summary, hoaxes tend to have more plain text than legitimate articles and
fewer links to external web pages outside of Wikipedia. This means that
non-hoax articles, in general, contain more references to links residing
outside Wikipedia. Such behavior is expected as a hoax's author would need to
put a significant effort into crafting external resources at which the hoax can
point.

\subsection{Traffic Analysis}

Recall that the cohort of a hoax is defined as all the non-hoax articles
created on the same day it was created. To understand the nature of the
relationship between the creation of hoaxes and the attention their respective
topics receive, we first seek to quantify the relative volume change before and
after this creation day. Here, a \emph{topic} is defined as all of the
(non-hoax) neighbors linked within the contents of an article i.e., its
(non-hoax) out-links. Traffic to Wikipedia is known to fluctuate following
circadian and weekly patterns, and is likely to depend on a host of additional,
unknown factors, such as the relative popularity of Wikipedia over the years,
the total number and geographic distribution of web
users~\cite{yasseri2012circadian}, etc. To account for these potential
confounding factors, \cite{ciampaglia2015production} proposed to quantify the
volume change in a way that controls for the circadian rhythm and the
fluctuating nature of traffic on the
Web~\cite{ciampaglia2015production,thompson1997wide}. They have shown that
studying traffic over a 14-day observation window, 7 days before and after the
creation day, considers both short spikes in attention and weekly changes in
traffic. The relative volume change is defined as:
\begin{align}
\frac{\Delta V}{V} & = \frac{V^{(b)} - V^{(a)}}{V^{(b)} + V^{(a)}}
\label{eq:delta_v} 
\end{align}
where $V^{(b)}$ and $V^{(a)}$ are respectively the median traffic to neighbors
in the 7 days before and after the creation of the article. According to
\cref{eq:delta_v}, $\Delta V / V > 0$ when the majority of traffic occurs
before an article is created, i.e., attention toward the topic of the articles
precedes its creation. When $\Delta V / V < 0$, attention tends instead to
follow the creation of the hoax. Note that our traffic data covers a period
spanning from December 2007 to August 2016. Since not all hoaxes in our dataset
fell within that time frame, $\Delta V / V$ was calculated only for the $83$
hoaxes (and their cohorts) whose creation dates fell within that period. 

\begin{figure}
  \includegraphics[width=\columnwidth]{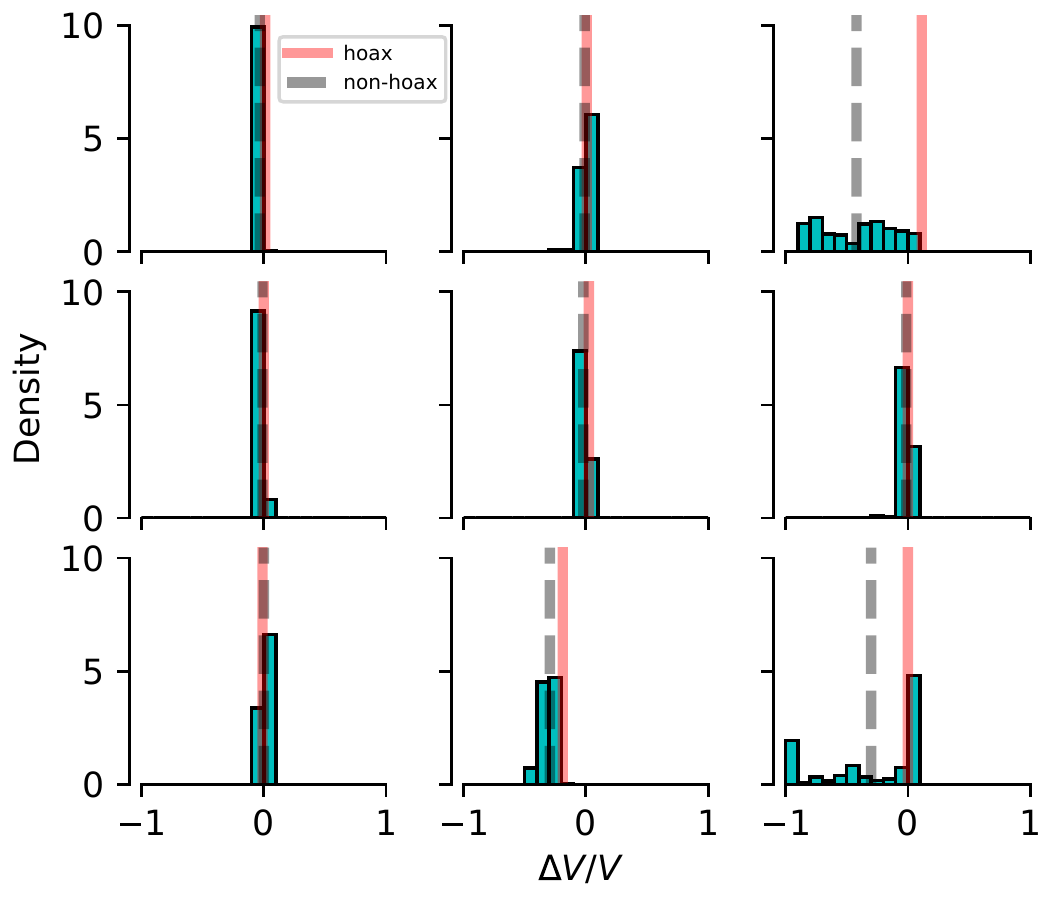}
  \caption{For a manual sample of hoaxes, the distribution of the $\Delta V' / V'$ values for each cohort (turquoise blue histograms) in comparison to the $\Delta V / V$ of the respective hoax (black dashed line). The $\Delta V / V$ of hoaxes tend to, in general, be higher than the mean of their cohorts (red solid line).}
  \label{fig:density_distributions}
\end{figure}

\Cref{fig:density_distributions} shows the distribution of the $\Delta V' / V'$
values for each cohort, the cohort mean, and the value of $\Delta V / V$ of the
corresponding hoax, for a manually selected sample of hoaxes collected from our
data. 

Having defined a way to quantify whether traffic to a given article preceded or
followed its creation, we want to determine whether hoaxes tend to have a
greater $\Delta V / V$ than legitimate articles in general. Unfortunately, we
know very little about the distribution of $\Delta V / V$ over multiple pages,
and how it has changed over the course of the history of Wikipedia. However, if
hoaxes do not differ from legitimate articles, then on average the difference
the $\Delta V/V$ of a hoax and that of its cohorts should be zero. Therefore,
we define: 
\begin{equation}
D = \frac{\Delta V}{V} - \mathsf{E}\left[\frac{\Delta V'}{V'}\right] =
\frac{\Delta V}{V} - \frac{1}{n}\sum_{i=1}^n \frac{\Delta V'_i} {V'_i}
\label{eq:D}
\end{equation}

where $\mathsf{E}\left[\frac{\Delta V'}{V'}\right]$ indicates the expected
$\Delta V' / V'$ of legitimate articles. Thus, when $D>0$ a hoax accumulates
more attention preceding its creation, compared to its cohort. 

To test whether $D > 0$ holds in general, we estimate the mean value of $D$ in
our sample of hoaxes, and used bootstrapping to compute the confidence interval
of the mean.
%
To perform bootstrapping, we resampled the original list of $D$ values $10,000$
times with replacement. 

\begin{figure*}
  \centering
  \includegraphics[width=\columnwidth]{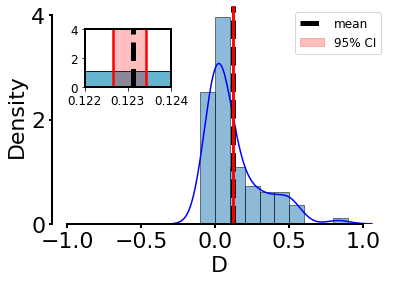}
  \includegraphics[width=\columnwidth]{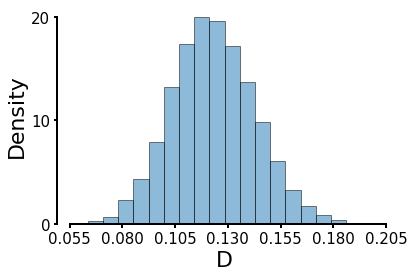}
  \caption{Left: Histogram of the relative traffic change differences $D$ (see \Cref{eq:D}). The black dashed line is the sample mean, and the red area the 95\% bootstrapped CI. The blue solid line is a kernel density estimate. The inset shows the sample mean relative to the confidence interval. Right: The sampling distribution of means obtained by bootstrapping $10,000$ samples with replacement.}
  \label{fig:bootstrap}
\end{figure*}

In general, we observe a trend in which hoaxes tend to have greater $\Delta V /
V$ than their cohort: $D > 0$ in 75 out of 83 of the hoaxes in our data. The
histogram in \cref{fig:bootstrap}~(left) shows the distribution of the
differences, and shows that the mean is approximately equal to $0.123$, with a
bootstrapped 95\% confidence interval of $(0.1227, 0.1234)$.

According to the Central Limit Theorem (CLT), the distribution of sample means
approximates the normal distribution with the increase of sample size $n$,
regardless of the original distribution of data~\cite{feller1991introduction}.
\Cref{fig:bootstrap}~(right) shows the distribution of the means for each of
the $10,000$ resampled vectors. It is worth noting that all of the means
returned were positive, implying a greater $\Delta V / V$ for the hoax. 

\section{Discussion and future work}
\label{sec:discussion_and_future_work}

Our study analyzes the role of online attention in the supply of disinformation
on Wikipedia (Q1). From an operational point of view, we study the relationship
between the creation of hoaxes and the traffic preceding each hoax's creation
day $d$ (Q2). To do so, we collected the view count of the out-link neighbors
of the hoaxes and their cohorts for $d\pm{}7$ days. Following prior
work~\cite{ciampaglia2015production}, to assess the allocation of attention
during that period, we calculated the relative traffic volume change, which
accounts for potential confounding factors due to traffic fluctuations. We
observe that 90\% of hoaxes have a higher $\Delta V / V$ than their respective
cohort and confirmed it by means of resampling. This indicates that, on
average, hoaxes tend to have more traffic accumulated before their creation
than after. In summary, our observed $D$ indicates that the generation of
hoaxes in Wikipedia is associated with prior consumption of information, in the
form of online attention, providing an answer to our original research question
(Q1).

This study has some limitations that need to be acknowledged. First of all, our
results are based on a list of only $83$ hoaxes. Even though we originally
collected a dataset that was twice the size of this one, we were limited by the
fact that not all hoaxes were covered in our traffic dataset. Future work
should extend our results to larger available samples of hoaxes (e.g.,
NPP-based) to ensure consistent results with prior work.

Additional limitations stem from our operational definition of the topic of a
new article (hoax or non-hoax). In this work, we relied on outgoing hyperlinks
(out-links) and neglected incoming hyperlinks (or in-links), owing to our lack
of access to data on hyperlinks to hoaxes. This data is present but not
publicly accessible in the Wikipedia database, presumably due internal
regulations within the Wikipedia project. In the future, we would like to
extend our analysis to include in-links as well. 

Future work should also consider a more advanced definition of an article's
topic that does not rely solely on hyperlinks, as they provide a very rough
notion of topic. Links to very generic entries like locations or dates (e.g.,
`United States of America' or `1492') typically convey little information about
an article's topic .

Third, our traffic dataset is based on an older definition of pagecounts, which
is affected by known limitations, including the presence of automated crawlers,
which are known to generate large amounts of hits in a short period of time. To
mitigate for the presence of this type of outliers, in our definitions of
traffic volume we rely on the median instead of the mean, which is more robust
to outliers. However, in the future we would like to include a more recent
traffic dataset that is not affected by this and other
biases~\cite{wikimedia2022research}.

In conclusion, our study sheds light on an important factor affecting the
supply of disinformation on the Web. Future work should extend our results to
venues other than Wikipedia, for example social media platforms like Facebook
or Twitter. In addition, other types of media (like video, audio, etc.) should
be considered --- hoaxes do not only come in the form of textual articles, and
attention is an effective incentive for people to keep spreading more
disinformation, regardless of its medium. Future work should also consider
studying the role of attention in versions of Wikipedia other than English. We
expect similar trends to ones observed here to apply to non-English language
editions as well. However, the signal may be weaker owing to lower traffic
volume of non-English language editions. A comparative analysis of the role of
attention in the supply of disinformation across cultures could shed more light
about these type of threats to the Web as a whole. 

\bibliography{acl2022}

\begin{thebibliography}{49}
\expandafter\ifx\csname natexlab\endcsname\relax\def\natexlab#1{#1}\fi

\bibitem[{5j9(2022)}]{5j92022github}
5j9. 2022.
\newblock Github inc., -- wikitextparser.
\newblock \url{https://github.com/5j9/wikitextparser}.
\newblock Last accessed: 09-March-2022.

\bibitem[{Adler et~al.(2010)Adler, de~Alfaro, and Pye}]{adler2010detecting}
B.~Thomas Adler, Luca de~Alfaro, and Ian Pye. 2010.
\newblock \href {http://ceur-ws.org/Vol-1176/CLEF2010wn-PAN-AdlerEt2010.pdf}
  {Detecting wikipedia vandalism using wikitrust - lab report for {PAN} at
  {CLEF} 2010}.
\newblock In \emph{{CLEF} 2010 LABs and Workshops, Notebook Papers, 22-23
  September 2010, Padua, Italy}, volume 1176 of \emph{{CEUR} Workshop
  Proceedings}, page n.p., Aachen, Germany. CEUR-WS.org.

\bibitem[{Allcott and Gentzkow(2017)}]{allcott2017social}
Hunt Allcott and Matthew Gentzkow. 2017.
\newblock \href {https://doi.org/10.1257/jep.31.2.211} {Social media and fake
  news in the 2016 election}.
\newblock \emph{Journal of Economic Perspectives}, 31(2):211--36.

\bibitem[{Amoruso et~al.(2020)Amoruso, Anello, Auletta, Cerulli, Ferraioli, and
  Raiconi}]{amoruso2020contrasting}
Marco Amoruso, Daniele Anello, Vincenzo Auletta, Raffaele Cerulli, Diodato
  Ferraioli, and Andrea Raiconi. 2020.
\newblock \href {https://doi.org/10.1613/jair.1.11509} {Contrasting the spread
  of misinformation in online social networks}.
\newblock \emph{Journal of Artificial Intelligence Research}, 69:847--879.

\bibitem[{Castillo et~al.(2011)Castillo, Mendoza, and
  Poblete}]{castillo2011information}
Carlos Castillo, Marcelo Mendoza, and Barbara Poblete. 2011.
\newblock \href {https://doi.org/10.1145/1963405.1963500} {Information
  credibility on twitter}.
\newblock In \emph{Proceedings of the 20th International Conference on World
  Wide Web}, WWW '11, pages 675--684, New York, NY, USA. Association for
  Computing Machinery.

\bibitem[{Ciampaglia et~al.(2015)Ciampaglia, Flammini, and
  Menczer}]{ciampaglia2015production}
Giovanni~Luca Ciampaglia, Alessandro Flammini, and Filippo Menczer. 2015.
\newblock \href {https://doi.org/10.1038/srep09452} {The production of
  information in the attention economy}.
\newblock \emph{Scientific Reports}, 5:9452.

\bibitem[{Dewey(2015)}]{dewey2015story}
Caitlin Dewey. 2015.
\newblock The story behind {Jar'Edo Wens}, the longest-running hoax in
  {Wikipedia}.
\newblock Last updated: 24-October-2018.

\bibitem[{Feller(1991)}]{feller1991introduction}
William Feller. 1991.
\newblock \emph{An Introduction to Probability Theory and Its Applications},
  2nd edition.
\newblock John Wiley \& Sons, Inc., New York, USA.

\bibitem[{García-Gavilanes et~al.(2017)García-Gavilanes, Mollgaard,
  Tsvetkova, and Yasseri}]{garcia2017memory}
Ruth García-Gavilanes, Anders Mollgaard, Milena Tsvetkova, and Taha Yasseri.
  2017.
\newblock \href {https://doi.org/10.1126/sciadv.1602368} {The memory remains:
  Understanding collective memory in the digital age}.
\newblock \emph{Science Advances}, 3(4):e1602368.

\bibitem[{Geiger and Halfaker(2013)}]{geiger2013levee}
R.~Stuart Geiger and Aaron Halfaker. 2013.
\newblock \href {https://doi.org/10.1145/2491055.2491061} {When the levee
  breaks: Without bots, what happens to {Wikipedia}'s quality control
  processes?}
\newblock In \emph{Proceedings of the 9th International Symposium on Open
  Collaboration}, WikiSym '13, New York, NY, USA. Association for Computing
  Machinery.

\bibitem[{Goldhaber(1997)}]{goldhaber1997attention}
Michael~H. Goldhaber. 1997.
\newblock \href {https://doi.org/10.5210/fm.v2i4.519} {The attention economy
  and the net}.
\newblock \emph{First Monday}, 2(4):n.p.

\bibitem[{Golovchenko et~al.(2020)Golovchenko, Buntain, Eady, Brown, and
  Tucker}]{golovchenko2020cross}
Yevgeniy Golovchenko, Cody Buntain, Gregory Eady, Megan~A. Brown, and Joshua~A.
  Tucker. 2020.
\newblock \href {https://doi.org/10.1177/1940161220912682} {Cross-platform
  state propaganda: {Russian} trolls on {Twitter} and {YouTube} during the 2016
  {U.S. Presidential Election}}.
\newblock \emph{The International Journal of Press/Politics}, 25(3):357--389.

\bibitem[{Grinberg et~al.(2019)Grinberg, Joseph, Friedland, Swire-Thompson, and
  Lazer}]{grinberg2019fake}
Nir Grinberg, Kenneth Joseph, Lisa Friedland, Briony Swire-Thompson, and David
  Lazer. 2019.
\newblock \href {https://doi.org/10.1126/science.aau2706} {Fake news on twitter
  during the 2016 u.s. presidential election}.
\newblock \emph{Science}, 363(6425):374--378.

\bibitem[{Guess et~al.(2019)Guess, Nagler, and Tucker}]{guess2019less}
Andrew Guess, Jonathan Nagler, and Joshua Tucker. 2019.
\newblock \href {https://doi.org/10.1126/sciadv.aau4586} {Less than you think:
  Prevalence and predictors of fake news dissemination on facebook}.
\newblock \emph{Science Advances}, 5(1):eaau4586.

\bibitem[{Guess et~al.(2020)Guess, Nyhan, and Reifler}]{guess2020exposure}
Andrew~M. Guess, Brendan Nyhan, and Jason Reifler. 2020.
\newblock \href {https://doi.org/10.1038/s41562-020-0833-x} {Exposure to
  untrustworthy websites in the 2016 us election}.
\newblock \emph{Nature Human Behaviour}, 4(5):472--480.

\bibitem[{Halfaker et~al.(2013)Halfaker, Geiger, Morgan, and
  Riedl}]{halfaker2013rise}
Aaron Halfaker, R.~Stuart Geiger, Jonathan~T. Morgan, and John Riedl. 2013.
\newblock \href {https://doi.org/10.1177/0002764212469365} {The rise and
  decline of an open collaboration system: How wikipedia{\textquoteright}s
  reaction to popularity is causing its decline}.
\newblock \emph{American Behavioral Scientist}, 57(5):664--688.

\bibitem[{Halfaker and Riedl(2012)}]{halfaker2012bots}
Aaron Halfaker and John Riedl. 2012.
\newblock \href {https://doi.org/10.1109/MC.2012.82} {Bots and cyborgs:
  {Wikipedia}'s immune system}.
\newblock \emph{IEEE Computer}, 45(3):79--82.

\bibitem[{Harpalani et~al.(2011)Harpalani, Hart, Singh, Johnson, and
  Choi}]{harpalani2011language}
Manoj Harpalani, Michael Hart, Sandesh Singh, Rob Johnson, and Yejin Choi.
  2011.
\newblock Language of vandalism: Improving {W}ikipedia vandalism detection via
  stylometric analysis.
\newblock In \emph{Proceedings of the 49th Annual Meeting of the Association
  for Computational Linguistics: Human Language Technologies}, pages 83--88,
  Portland, Oregon, USA. Association for Computational Linguistics.

\bibitem[{Iglewicz and Hoaglin(1993)}]{iglewicz1993how}
Boris Iglewicz and David~C. Hoaglin. 1993.
\newblock How to detect and handle outliers.
\newblock In Edward~F. Mykytka, editor, \emph{The ASQC Basic References in
  Quality Control: Statistical Techniques}, volume~16. ASQC, Milwaukee, WI,
  USA.

\bibitem[{King et~al.(2017)King, Pan, and Roberts}]{king2017chinese}
Gary King, Jennifer Pan, and Margaret~E. Roberts. 2017.
\newblock \href {https://doi.org/10.1017/S0003055417000144} {How the chinese
  government fabricates social media posts for strategic distraction, not
  engaged argument}.
\newblock \emph{American Political Science Review}, 111(3):484--501.

\bibitem[{Kumar et~al.(2016)Kumar, West, and
  Leskovec}]{kumar2016disinformation}
Srijan Kumar, Robert West, and Jure Leskovec. 2016.
\newblock \href {https://doi.org/10.1145/2872427.2883085} {Disinformation on
  the web: Impact, characteristics, and detection of wikipedia hoaxes}.
\newblock In \emph{Proceedings of the 25th International Conference on World
  Wide Web}, WWW '16, page 591–602, Republic and Canton of Geneva, CHE.
  International World Wide Web Conferences Steering Committee.

\bibitem[{Lazer et~al.(2018)Lazer, Baum, Benkler, Berinsky, Greenhill, Menczer,
  Metzger, Nyhan, Pennycook, Rothschild, Schudson, Sloman, Sunstein, Thorson,
  Watts, and Zittrain}]{lazer2018science}
David M.~J. Lazer, Matthew~A. Baum, Yochai Benkler, Adam~J. Berinsky, Kelly~M.
  Greenhill, Filippo Menczer, Miriam~J. Metzger, Brendan Nyhan, Gordon
  Pennycook, David Rothschild, Michael Schudson, Steven~A. Sloman, Cass~R.
  Sunstein, Emily~A. Thorson, Duncan~J. Watts, and Jonathan~L. Zittrain. 2018.
\newblock \href {https://doi.org/10.1126/science.aao2998} {The science of fake
  news}.
\newblock \emph{Science}, 359(6380):1094--1096.

\bibitem[{{MediaWiki contributors}(2021)}]{mediawiki2022text}
{MediaWiki contributors}. 2021.
\newblock Extension:textextracts --- mediawiki.
\newblock
  \url{https://www.mediawiki.org/w/index.php?title=Extension:TextExtracts&oldid=4940004}.
\newblock Last accessed: 9-March-2022.

\bibitem[{{MediaWiki contributors}(2022{\natexlab{a}})}]{mediawiki2022main}
{MediaWiki contributors}. 2022{\natexlab{a}}.
\newblock Api:main page --- mediawiki.
\newblock
  \url{https://www.mediawiki.org/w/index.php?title=API:Main_page&oldid=5019333}.
\newblock Last accessed: 9-March-2022.

\bibitem[{{MediaWiki
  contributors}(2022{\natexlab{b}})}]{mediawiki2022revisions}
{MediaWiki contributors}. 2022{\natexlab{b}}.
\newblock Api:revisions --- mediawiki.
\newblock
  \url{https://www.mediawiki.org/w/index.php?title=API:Revisions&oldid=5037632}.
\newblock Last accessed: 9-March-2022.

\bibitem[{Moon(2022)}]{moon2022chinese}
Mariella Moon. 2022.
\newblock \href
  {https://www.engadget.com/chinese-wikipedia-editor-fake-russian-medieval-history-122001604.html}
  {A {Chinese} {Wikipedia} editor spent years writing fake {Russian} medieval
  history}.
\newblock Last accessed: 13-Sep-2022.

\bibitem[{Okoli et~al.(2014)Okoli, Mehdi, Mesgari, Nielsen, and
  Lanam{\"{a}}ki}]{okoli2014wikipedia}
Chitu Okoli, Mohamad Mehdi, Mostafa Mesgari, Finn~{{\AA}}rup Nielsen, and Arto
  Lanam{\"{a}}ki. 2014.
\newblock \href {https://doi.org/10.1002/asi.23162} {Wikipedia in the eyes of
  its beholders: A systematic review of scholarly research on wikipedia readers
  and readership}.
\newblock \emph{Journal of the Association for Information Science and
  Technology}, 65(12):2381--2403.

\bibitem[{Potthast et~al.(2008)Potthast, Stein, and
  Gerling}]{potthast2008automatic}
Martin Potthast, Benno Stein, and Robert Gerling. 2008.
\newblock Automatic vandalism detection in wikipedia.
\newblock In \emph{Advances in Information Retrieval}, pages 663--668, Berlin,
  Heidelberg. Springer Berlin Heidelberg.

\bibitem[{S{\'{a}}ez{-}Trumper(2019)}]{diego2019online}
Diego S{\'{a}}ez{-}Trumper. 2019.
\newblock \href {http://arxiv.org/abs/1910.12596} {Online disinformation and
  the role of wikipedia}.
\newblock \emph{CoRR}, abs/1910.12596.

\bibitem[{{Similarweb LTD}(2022)}]{similarweb2022websites}
{Similarweb LTD}. 2022.
\newblock Top websites ranking.
\newblock \url{https://www.similarweb.com/top-websites/}.
\newblock Last accessed: Mar-09-2022.

\bibitem[{Simon(1971)}]{simon1971designing}
Herbert~A Simon. 1971.
\newblock Designing organizations for an information-rich world.
\newblock In Martin Greenberger, editor, \emph{Computers, communications, and
  the public interest}, volume~72, pages 37--52. Johns Hopkins Press,
  Baltimore.

\bibitem[{Smets et~al.(2008)Smets, Goethals, and Verdonk}]{smets2008automatic}
Koen Smets, Bart Goethals, and Brigitte Verdonk. 2008.
\newblock Automatic vandalism detection in wikipedia: Towards a machine
  learning approach.
\newblock In \emph{Proceedings of the 2008 {AAAI} Workshop on {Wikipedia} and
  Artificial Intelligence: An Evolving Synergy}, pages 43--48, Palo Alto, CA,
  USA. AAAI.

\bibitem[{Starbird(2017)}]{starbird2017examining}
Kate Starbird. 2017.
\newblock \href {https://aaai.org/ocs/index.php/ICWSM/ICWSM17/paper/view/15603}
  {Examining the alternative media ecosystem through the production of
  alternative narratives of mass shooting events on twitter}.
\newblock In \emph{Proc. of the International AAAI Conference on Web and Social
  Media}, pages 230--239, Palo Alto, CA, USA. AAAI.

\bibitem[{Thompson et~al.(1997)Thompson, Miller, and Wilder}]{thompson1997wide}
K.~Thompson, G.J. Miller, and R.~Wilder. 1997.
\newblock \href {https://doi.org/10.1109/65.642356} {Wide-area internet traffic
  patterns and characteristics}.
\newblock \emph{IEEE Network}, 11(6):10--23.

\bibitem[{Vi\'{e}gas et~al.(2004)Vi\'{e}gas, Wattenberg, and
  Dave}]{viegas2004studying}
Fernanda~B. Vi\'{e}gas, Martin Wattenberg, and Kushal Dave. 2004.
\newblock \href {https://doi.org/10.1145/985692.985765} {Studying cooperation
  and conflict between authors with {History Flow} visualizations}.
\newblock In \emph{Proceedings of the SIGCHI Conference on Human Factors in
  Computing Systems}, CHI '04, pages 575--582, New York, NY, USA. Association
  for Computing Machinery.

\bibitem[{Wardle and Derakhshan(2017)}]{wardle2017information}
Claire Wardle and Hossein Derakhshan. 2017.
\newblock Information disorder: toward an interdisciplinary frameworkfor
  research and policy making.
\newblock Technical Report DGI(2017)09, Council of Europe, Strasbourg, FR.

\bibitem[{{Wikimedia Foundation, Inc.}(2022{\natexlab{a}})}]{wikimedia2022help}
{Wikimedia Foundation, Inc.} 2022{\natexlab{a}}.
\newblock Help: Page name.
\newblock \url{https://meta.wikimedia.org/wiki/Help:Page_name}.
\newblock Last accessed: 09-March-2022.

\bibitem[{{Wikimedia Foundation, Inc.}(2022{\natexlab{b}})}]{wikimedia2022page}
{Wikimedia Foundation, Inc.} 2022{\natexlab{b}}.
\newblock Page view statistics for {Wikimedia} projects.
\newblock \url{https://dumps.wikimedia.org/other/pagecounts-raw/}.
\newblock Last accessed: 09-March-2022.

\bibitem[{{Wikimedia Foundation,
  Inc.}(2022{\natexlab{c}})}]{wikimedia2022research}
{Wikimedia Foundation, Inc.} 2022{\natexlab{c}}.
\newblock Research:page\_view.
\newblock \url{https://meta.wikimedia.org/wiki/Research:Page\_view}.
\newblock Last accessed: 13-September -2022.

\bibitem[{{Wikimedia Foundation,
  Inc.}(2022{\natexlab{d}})}]{wikimedia2022wikimedia}
{Wikimedia Foundation, Inc.} 2022{\natexlab{d}}.
\newblock {Wikimedia Statistics -- English Wikipedia}.
\newblock \url{https://stats.wikimedia.org/\#/en.wikipedia.org}.
\newblock Last accessed 09-March-2022.

\bibitem[{{Wikipedia contributors}(2021)}]{wikimedia2022vandalism}
{Wikipedia contributors}. 2021.
\newblock Wikipedia:vandalism does not matter.
\newblock
  \url{https://en.wikipedia.org/wiki/Wikipedia:Vandalism_does_not_matter}.
\newblock Last accessed: Mar-09-2022.

\bibitem[{{Wikipedia contributors}(2022{\natexlab{a}})}]{wikimedia2022list}
{Wikipedia contributors}. 2022{\natexlab{a}}.
\newblock Wikipedia:list of hoaxes on wikipedia.
\newblock
  \url{https://en.wikipedia.org/wiki/Wikipedia:List_of_hoaxes_on_Wikipedia}.
\newblock Last accessed: Mar-09-2022.

\bibitem[{{Wikipedia contributors}(2022{\natexlab{b}})}]{wikimedia2022new}
{Wikipedia contributors}. 2022{\natexlab{b}}.
\newblock Wikipedia:new pages patrol.
\newblock \url{https://en.wikipedia.org/wiki/Wikipedia:New_pages_patrol}.
\newblock Last accessed: Mar-09-2022.

\bibitem[{{Wikipedia contributors}(2022{\natexlab{c}})}]{wikimedia2022size}
{Wikipedia contributors}. 2022{\natexlab{c}}.
\newblock Wikipedia:size of wikipedia.
\newblock \url{https://en.wikipedia.org/wiki/Wikipedia:Size_of_Wikipedia}.
\newblock Last accessed: Mar-09-2022.

\bibitem[{{Wikitech contributors}(2021)}]{wikitech2021toolforge}
{Wikitech contributors}. 2021.
\newblock Portal:toolforge --- wikitech.
\newblock
  \url{https://wikitech.wikimedia.org/w/index.php?title=Portal:Toolforge&oldid=1932575}.
\newblock Last accessed: 9-March-2022.

\bibitem[{Yasseri and Menczer(2021)}]{yasseri2021crowdsourcing}
Taha Yasseri and Filippo Menczer. 2021.
\newblock \href {https://doi.org/10.48550/ARXIV.2104.13754} {Can crowdsourcing
  rescue the social marketplace of ideas?}
\newblock Technical report, arXiv.

\bibitem[{Yasseri et~al.(2012)Yasseri, Sumi, and
  Kertész}]{yasseri2012circadian}
Taha Yasseri, Robert Sumi, and János Kertész. 2012.
\newblock \href {https://doi.org/10.1371/journal.pone.0030091} {Circadian
  patterns of wikipedia editorial activity: A demographic analysis}.
\newblock \emph{PLOS ONE}, 7(1):1--8.

\bibitem[{Zannettou et~al.(2019)Zannettou, Caulfield, De~Cristofaro,
  Sirivianos, Stringhini, and Blackburn}]{zannettou2019disinformation}
Savvas Zannettou, Tristan Caulfield, Emiliano De~Cristofaro, Michael
  Sirivianos, Gianluca Stringhini, and Jeremy Blackburn. 2019.
\newblock \href {https://doi.org/10.1145/3308560.3316495} {Disinformation
  warfare: Understanding state-sponsored trolls on twitter and their influence
  on the web}.
\newblock In \emph{Companion Proceedings of The 2019 World Wide Web
  Conference}, WWW '19, pages 218--226, New York, NY, USA. Association for
  Computing Machinery.

\bibitem[{Zareie and Sakellariou(2021)}]{zareie2021minimizing}
Ahmad Zareie and Rizos Sakellariou. 2021.
\newblock \href {https://doi.org/10.1016/j.jnca.2021.103094} {Minimizing the
  spread of misinformation in online social networks: A survey}.
\newblock \emph{Journal of Network and Computer Applications}, 186:103094.

\end{thebibliography}
\bibliographystyle{acl_natbib}

\end{document}